\documentclass[twocolumn]{aastex62}

\submitjournal{ApJL}

\shorttitle{The warmer climate under high obliquity}
\shortauthors{Kang}

\begin{document}

\title{Mechanisms leading to a warmer climate on high obliquity planets}

\correspondingauthor{Wanying Kang}
\email{wanyingkang@g.harvard.edu}

\author[0000-0002-4615-3702]{Wanying Kang}
\affil{School of Engineering and Applied Sciences\\
  Harvard University\\
  Cambridge, MA 02138, USA}

\begin{abstract}
  A consistent finding of high obliquity simulations is that they are warmer than their low obliquity counterparts when the climate is cold. Ice-albedo feedback has been suggested as a possible mechanism. In this study, we find that warmer climate under high obliquity holds with varying insolations, including almost ice-free conditions. We try to understand the mechanisms through a series of feedback suppression experiments. Turning off the ice-albedo feedback, the temperature contrast between high and low obliquity remains significant, but it vanishes when the cloud radiation effects or the seasonal variation is turned off. This suggests the warmer climate on high obliquity planets does not rely completely on the existence of ice, and therefore holds at high insolation. In that regime, the surface temperature, and hence the cloud formation, lags behind the substellar point, leading to inefficient sunlight reflection and warmer climate.
\end{abstract}


\section{Introduction} \label{sec:intro}

An exoplanet may have a large obliquity or large obliquity variability depending on the initial angular momentum of the nebulae that formed that planet, continental movement \citep{Williams-Kasting-Frakes-1998:low}, gravitational interference from other bodies \citep{Correia-Laskar-2010:tidal}, and the history of its orbital migration \citep{Brunini-2006:origin}. In our solar system, for example, Mars's obliquity chaotically varies from 0 to 60 degrees \citep{Laskar-Robutel-1993:chaotic}, and Venus and Uranus have obliquities close to 180 and 90 degrees respectively \citep{Carpenter-1966:study}. Among exoplanets, high obliquity planets are expected to widely exist in the universe due to secular resonance-driven spin-orbit coupling \citep{Millholland-Laughlin-2019:obliquity}. The earth may also have been in a high obliquity state in the past, and the high obliquity scenario has been used to explain the ``Faint-Young Sun paradox'' and the two low-latitude glacial events in Early and Late Proterozoic \citep{Jenkins-2000:global, Jenkins-2001:high, Jenkins-2003:gcm}.

Planets with extremely high obliquity have been shown to be completely or partially ice-free at a much farther distance from the host star using a 3D general circulation model \citep[GCM,][]{Linsenmeier-Pascale-Lucarini-2015:climate, Kilic-Raible-Stocker-2017:multiple, Kilic-Lunkeit-Raible-et-al-2018:stable} and a conceptual energy balance model \citep[EBM,][]{Rose-Cronin-Bitz-2017:ice, Armstrong-Barnes-Domagal-Goldman-et-al-2014:effects}. 
\citet{Jenkins-2000:global} showed that the climate could be warmer due to a larger obliquity, providing a potential explanation for the warm climate during the Earth's early history, in spite of a 20-30\% dimmer sun.
Even in less extreme obliquity variations, as is the case with present-day Earth (obliquity fluctuates between 22 and 24.5 degrees), terminations of glaciation have been shown to be linked to the high obliquity periods \citep{Paillard-2001:glacial, Paillard-1998:pleistocene, Huybers-Wunsch-2005:obliquity} during the Pleistocene. 
This raises the question: Why is the climate warmer on high obliquity planets?


Given that high obliquity planets tend to have low ice coverage \citep{Linsenmeier-Pascale-Lucarini-2015:climate, Kilic-Lunkeit-Raible-et-al-2018:stable}, it has been suggested that the ice-albedo feedback is the cause of the warmer climate under high obliquity. However, even aquanplanet simulations suggest that the global surface temperature rises with obliquity, with concomitant reduction in cloud coverage and sunlight reflection explaining the warming \citep{Nowajewski-Rojas-Rojo-et-al-2018:atmospheric}. When small changes of obliquity are applied, cloud feedback and lapse rate feedback have also been found to significantly contribute to the warmer climate \citep{Mantsis-Clement-Broccoli-et-al-2011:climate}. Even for tidally-locked planets, high obliquity leads to a warmer climate because of the low cloud coverage in the day side \citep{Wang-Liu-Tian-et-al-2016:effects}. All of the above suggest that the cloud feedback, ice-albedo feedback, and lapse rate feedback may lead to a warmer climate under high obliquity in general. However, it still needs to be investigated whether the relative warmness is a universal phenomenon regardless of other parameters and whether it would hold even without clouds. If the clouds do tend to reflect less under high obliquity, then we need to understand what mechanisms underlie this.

In this study, we show that the relatively warm climate under high obliquity is valid in a wide range of insolation (consistent with \citet{Nowajewski-Rojas-Rojo-et-al-2018:atmospheric}), independent of the existence of sea ice. We then explore the mechanisms using a series of feedback suppression experiments, sequentially turning off ice-albedo feedback, cloud radiation effects and seasonal cycle. 

\section{Methods}
\label{sec:methods}

The model used here is Community Earth System Model version 1.2.1 \cite[CESM,][]{Neale-Chen-Gettelman-2010:description}, modified by \citet[][code are available on GitHub\footnote{ https://github.com/storyofthewolf/ExoRT and https://github.com/storyofthewolf/ExoCAM}]{Kopparapu-Wolf-Arney-et-al-2017:habitable} to include the following two features: 1) increased spectral resolution in the near infrared for a more realistic radiation calculation, and 2) more frequent sub-step dynamic adjustment to improve numerical stability. Thanks to the fine spectral resolution, this radiation scheme was shown to be more robust at the high temperature end, while the default CESM radiative transfer model underestimates both longwave and shortwave water vapor absorption \citep{Yang-Leconte-Wolf-et-al-2016:differences}. We consider H$_2$O as the only greenhouse gas in the atmosphere for simplicity, while ignoring any CO$_2$ absorption. This makes it easier to understand the surface temperature difference between two experiments. 
The atmospheric circulation is simulated by a finite-volume dynamical core, with approximately 1.9 degree horizontal resolution and 40 vertical layers extending to 0.8 mb. This atmospheric model is coupled with a 50 meter deep slab ocean. Horizontal ocean heat transport is not included for simplicity, since it has been shown to play a minor role in the surface temperature, compared to a large change in obliquity \citep{Jenkins-2003:gcm}. We choose to explicitly simulate sea ice formation using the Community Ice CodE (CICE) version 4, rather than use an aquaplanet setup, in order to capture the full climate response.

We first examine whether the warming at high obliquity holds for different insolations. For both 80 deg obliquity and 0 deg obliquity, we gradually increase the insolation from 1365 W/m$^2$ to 1750 W/m$^2$ during a 100 year simulation, with the expectation that the transient insolation change is slow enough to allow the atmosphere to adjust to its equilibrium state. Both experiments are initialized from the snowball state (i.e., 100\% ice coverage) and are run to equilibrium before we start to increase the insolation. Both cases can be stably integrated at 1750 W/m$^2$ for at least 50 years. Adding another 50 W/m$^2$ leads to runaway greenhouse (model crashes).

We then perform a series of feedback suppression experiments to understand the mechanisms that cause the warming under high obliquity. We start with the control experiments including all climate feedbacks, then we turn off the ice-albedo feedback, and finally turn off the cloud radiation effects in one branch, and the seasonal cycle in the other branch. All the feedback suppression experiments apply a 1365 W/m$^2$ insolation, zero eccentricity, and 1 bar N$_2$ atmosphere, and are run under both 0 deg obliquity and 80 deg obliquity. In the control experiments, the ocean has 0.06 albedo for radiation at all frequencies, while sea ice and snow have much higher albedo of 0.67 (0.3) and 0.8 (0.68) for the visible (infrared) radiation respectively. We then turn off the ice-albedo feedback by applying the ocean albedo (0.06), everywhere on the globe. After that, cloud radiation effects and the seasonal cycle are turned off individually. To shut down the cloud radiation effects, we set the cloud optical thickness to zero. To turn off the seasonal cycle, annual mean insolation is applied to each latitude all year round.

One should keep in mind that, in the climate system where all fields are affecting and are affected by each other, causality could be ambiguous. By turning off a certain mechanism, we also exclude the compensating effects and indirect response of other processes \cite[][]{Cai-Lu-2009:new}, and thus, we do not expect the climate response to different feedbacks to be additive.

\section{Results}
\label{sec:results}

\begin{figure}[htp!]
 \centering
\includegraphics[width=0.5\textwidth]{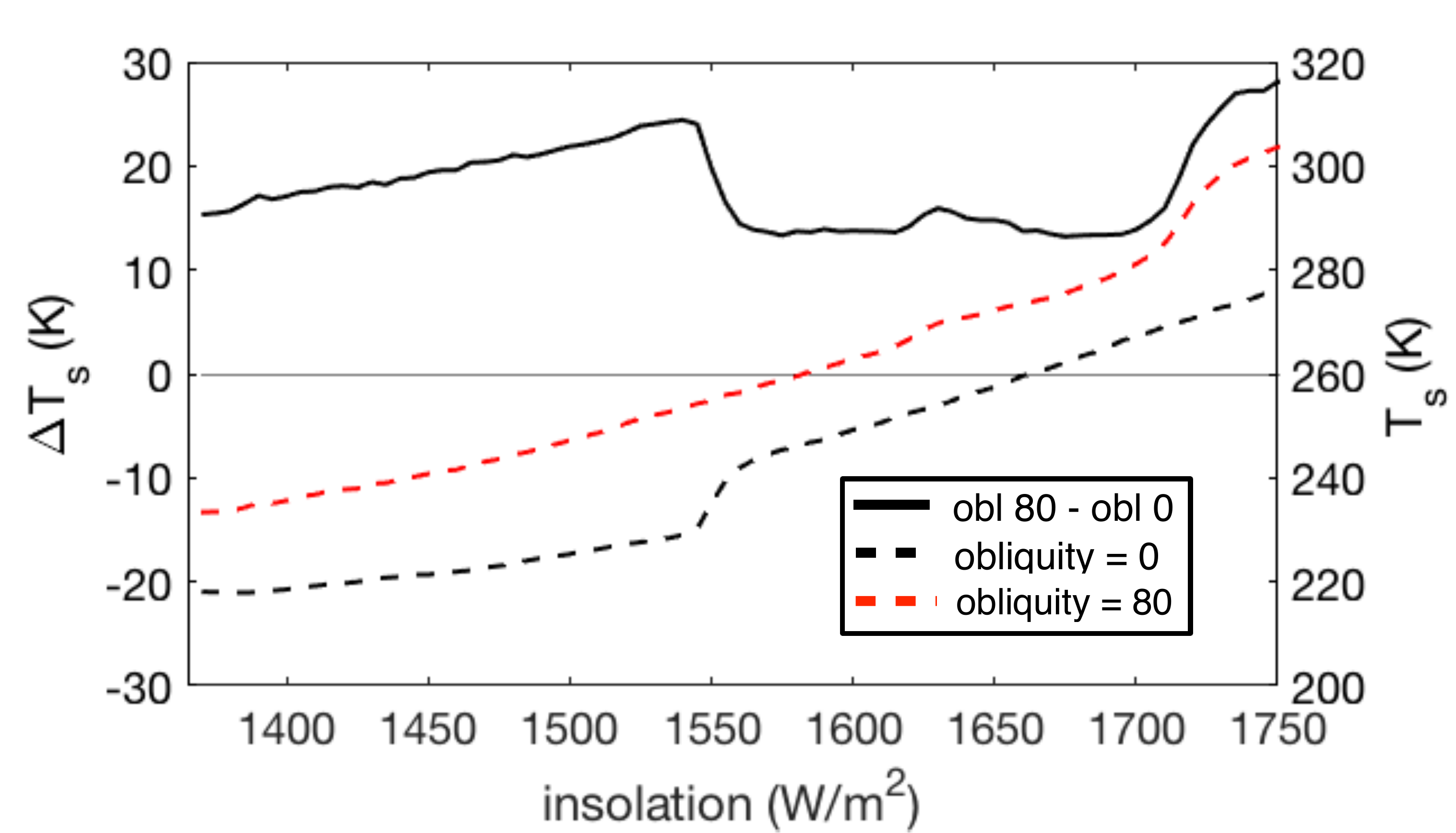}
 \caption{Global annual mean surface temperature difference between the high and low obliquity, as insolation gradually increases (solid curve, left axis). Shown on the right axis are the high obliquity (dashed red) and low obliquity (dashed black) mean surface temperature.}
 \label{fig:Ts-insolation}
\end{figure}

We first examine whether the relative warmness at high obliquity holds despite varying insolation. Shown in Fig.~\ref{fig:Ts-insolation} are the global annual mean surface temperature progressions in the low obliquity experiment (black dash) and in the high obliquity experiment (red dash) as insolation gradually increases. The high obliquity climate is always warmer than the low obliquity climate by over 10 K. This holds even when approaching the runaway greenhouse state, suggesting that the warming at high obliquity holds not only when the climate is cold, as studied in \citet{Jenkins-2000:global} and \citet{Linsenmeier-Pascale-Lucarini-2015:climate}, but also when the climate is warm enough to be almost ice-free, consistent with \citet{Nowajewski-Rojas-Rojo-et-al-2018:atmospheric}. There is a temperature jump in the zero obliquity experiment at around 1550 W/m$^2$, marking an abrupt transition out of the snowball state. A similar jump is seen in the high obliquity experiment around 1700 W/m$^2$, and this is caused by an abrupt transition into a complete ice-free state.

\begin{figure*}[htp!]
 \centering
\includegraphics[width=\textwidth]{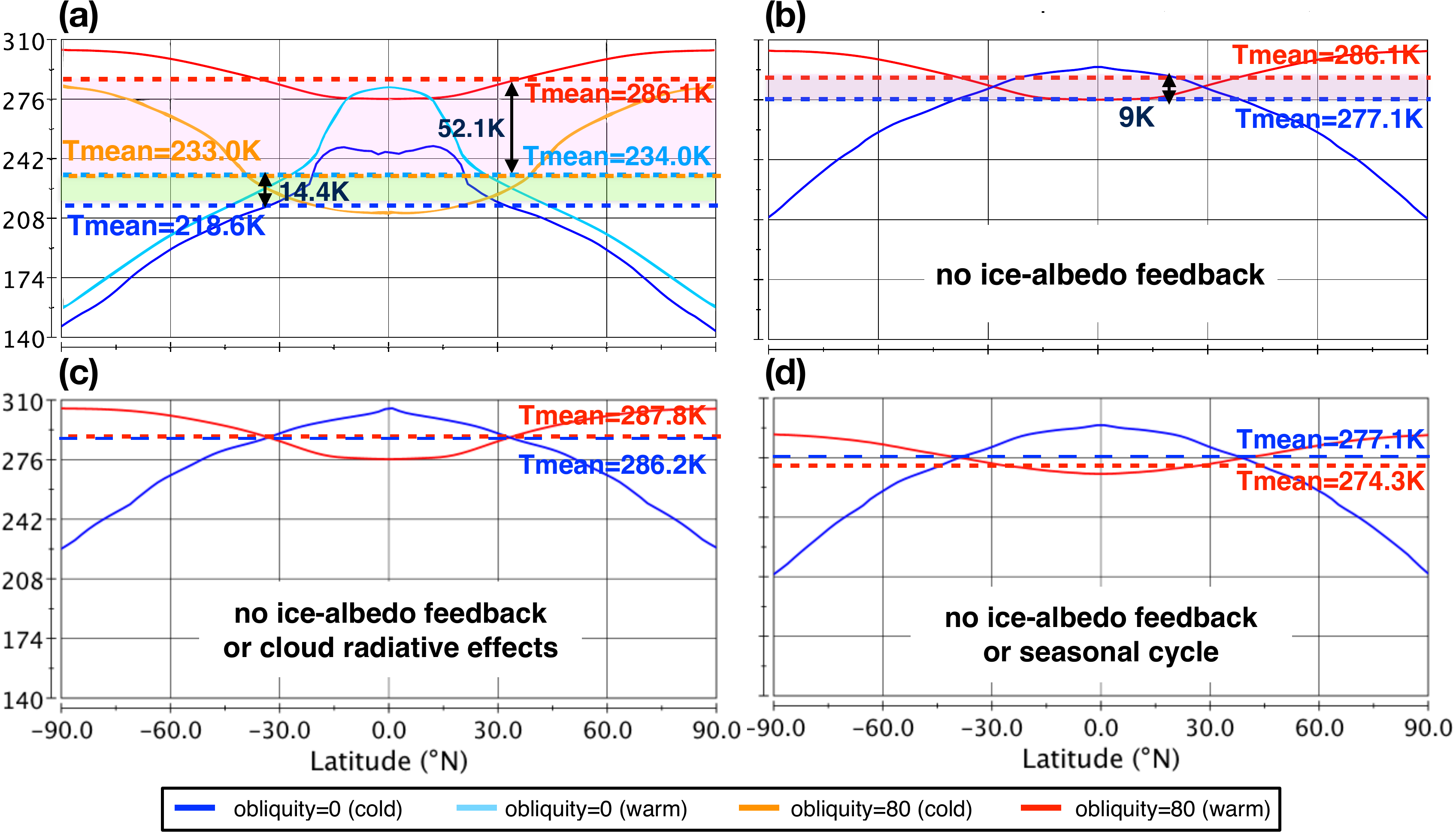}
 \caption{Annual mean latitudinal surface temperature profile under low obliquity (blue) and high obliquity (red). Global mean annual mean surface temperature is marked by dashed curve for all cases, and the difference is highlighted by purple shadings. Shown are for (a) control experiments with all feedbacks on, (b) experiments without ice-albedo feedback, (c) experiments without ice-albedo feedback or cloud radiation effects, and (d) experiments without ice-albedo feedback or seasonal cycle. In (a), there are two equilibrium states for both of the high and low obliquity climate. Dark blue and orange denote the colder equilibrium states, and light blue and red denote the warmer states. In (b,c,d), there is only one equilibrium state for high and low obliquity climate, and they are plotted in dark blue and red curves.}
 \label{fig:Ts-lat-feedbacks}
\end{figure*}

We then try to understand what mechanisms account for the relative warmness under high obliquity using 1365 W/m$^2$ as an example. Due to the positive ice-albedo feedback, we find two equilibrium states for both high and low obliquity experiments by initiating the model from either a snowball state or an aquaplanet state  \citep[see][]{Kilic-Raible-Stocker-2017:multiple, Linsenmeier-Pascale-Lucarini-2015:climate}. One equilibrium state has greater ice coverage, resulting in a higher albedo and thus a colder climate, while the other equilibrium state has less ice and thus a warmer climate. The latitudinal profile of annual-mean surface temperature is shown in Fig.~\ref{fig:Ts-lat-feedbacks}a for the four states. Comparing the warm branches of the high and low obliquity yields a $286.1-234.0=52.1$ K difference (high obliquity minus low obliquity). The high obliquity simulation is completely ice free, while the zero obliquity freezes to 14N/S. The strong contrast in the ice coverage between the two simulations accounts for the large temperature contrast. In the cold branch, the temperature contrast is smaller, $233.0-218.6=14.4$ K (consistent with Fig.~\ref{fig:Ts-insolation}). The high obliquity experiment shows strong seasonal variation of ice coverage, while the low obliquity experiment is almost in a snowball state since CO$_2$ is not considered. Warming under high obliquity has also been noted by \citet{Jenkins-2000:global} and \citet{Nowajewski-Rojas-Rojo-et-al-2018:atmospheric}. In \citet{Jenkins-2000:global}, a temperature difference greater than 60 K has been found between the high and low obliquity simulations, when other parameters are fixed, as seen in the warm branch of our experiments.

High obliquity planets tend to have less ice coverage, as also noted by \citet{Linsenmeier-Pascale-Lucarini-2015:climate} and \citet{Kilic-Lunkeit-Raible-et-al-2018:stable}. In the high obliquity cold branch, long and direct sunlight during the polar day would melt the ice at high latitudes ($>$45N/S here). In the high obliquity warm branch, where the ice coverage is low and the shortwave absorption is enhanced, the ocean remains ice-free year round. The suppression of ice formation during polar night is accomplished by the heat stored in the ocean from the previous summer. Coupled atmosphere-ocean modeling has shown that an ocean can keep the high latitudes warmer than the equator all year around under high obliquity \citep{Ferreira-Marshall-OGorman-et-al-2014:climate}. The lower ice coverage under high obliquity therefore leads to a smaller albedo and a warmer climate.

To evaluate the role played by ice albedo feedback, we switch it off (see section~\ref{sec:methods} for details). The results are shown in Fig.~\ref{fig:Ts-lat-feedbacks}(b). The temperature contrast between the high and low obliquity simulations reduces to 9 K\footnote{Since feedback suppression method does not guarantee additivity, this result does not mean that ice reflection contributes the remaining 40 K temperature difference.}. On the one hand, this indicates that most of the warming at high obliquity we see in the control experiments can be attributed to the ice-albedo feedback, as suggested by \citet{Linsenmeier-Pascale-Lucarini-2015:climate} and \citet{Kilic-Lunkeit-Raible-et-al-2018:stable}. This is the case because the climate is cold enough to allow ice-albedo feedback to dominate other feedbacks, like cloud and water vapor feedbacks. On the other hand, this also suggests that high obliquity planets tend to be warmer than the low obliquity equivalents even without ice. Similar results have also been found in \citet{Nowajewski-Rojas-Rojo-et-al-2018:atmospheric}, where the authors altered the obliquity in a warm climate that is naturally ice-free.

\begin{figure*}[htp!]
 \centering
\includegraphics[width=\textwidth]{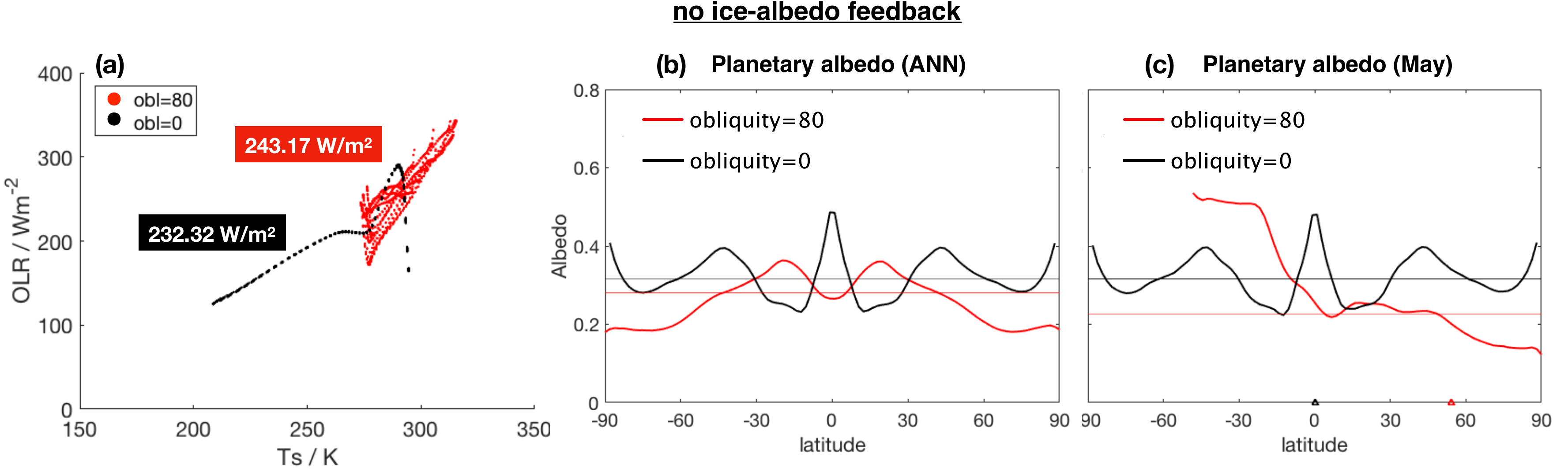}
 \caption{The mechanism that causes the high obliquity to be warmer than the low obliquity even without the ice-albedo feedback. (a) Scatter plots of OLR (outgoing longwave radiation) against surface temperature for the experiments without ice-albedo feedback. Each dot corresponds to one monthly climatology at one latitude. Global mean annual mean OLR is marked in text box. This should be in balance with the total absorbed solar radiation. (b,c) Planetary albedo as a function of latitude for (thick red) high obliquity and (thick black) low obliquity, both without ice-albedo feedback. Global area averages are plotted in thin curves. Shown are for (b) annual mean and (c) May climatology. In panel (b), substellar point are marked by red and black triangles on the x-axis. The albedo is not shown for the places receiving no solar radiation. Global mean albedo is evaluated by taking the ratio between the global reflected shortwave and the global incoming shortwave. }
 \label{fig:OLR-albedo}
\end{figure*}

There are (at least) two possible explanations for the remaining 9K difference: weaker reflection or stronger greenhouse effect. As suggested by the heat budget in \citet{Nowajewski-Rojas-Rojo-et-al-2018:atmospheric}, the sunlight reflection gets weaker under higher obliquity. The same conclusion holds here. We distinguish the two possibilities by checking whether the outgoing longwave radiation (OLR) increases proportionally with the surface temperature. If so, the ``weaker reflection'' explanation seems more relevant, and a stronger absorbed solar radiation is required for energy balance. Shown in Fig.~\ref{fig:OLR-albedo}(a) is the scatter plot of OLR against surface temperature for the ice-albedo feedback suppressed experiments, with each dot representing one latitude and one month's climatology. The dots of the zero obliquity experiment (black) align with those of the high obliquity experiment, consistent with \citet{Koll-Cronin-2018:earths}. Together they roughly follow one linear relationship. A higher mean surface temperature in the high obliquity experiment corresponds to a higher OLR, which has to be in balance with a stronger absorption (or less reflection) of solar radiation. The global annual mean OLR in the high obliquity experiment is 243.17 W/m$^2$, about 11 W/m$^2$ higher than that in the zero obliquity experiment. This indicates that, compared to a zero obliquity planet, a high obliquity planet tends to reflect less incoming solar radiation, ending up with higher mean surface temperature, even without ice/snow.

We then look into the spatial and temporal distribution of albedo. Shown in Fig.~\ref{fig:OLR-albedo}(b,c) are the latitude dependencies of planetary albedo for both high and low obliquity experiments without ice-albedo feedback. Global mean planetary albedo is marked by thin straight lines. On average, the high obliquity planet has a lower albedo, as expected. In particular, the high latitudes, which receive the most sunlight, have a quite low albedo of 0.2. In contrast, the zero obliquity planet reflects strongest at the equator where the insolation also peaks. Without sea ice reflection, the dominant factor for albedo becomes the cloud distribution.

\begin{figure*}[htp!]
 \centering \includegraphics[width=\textwidth]{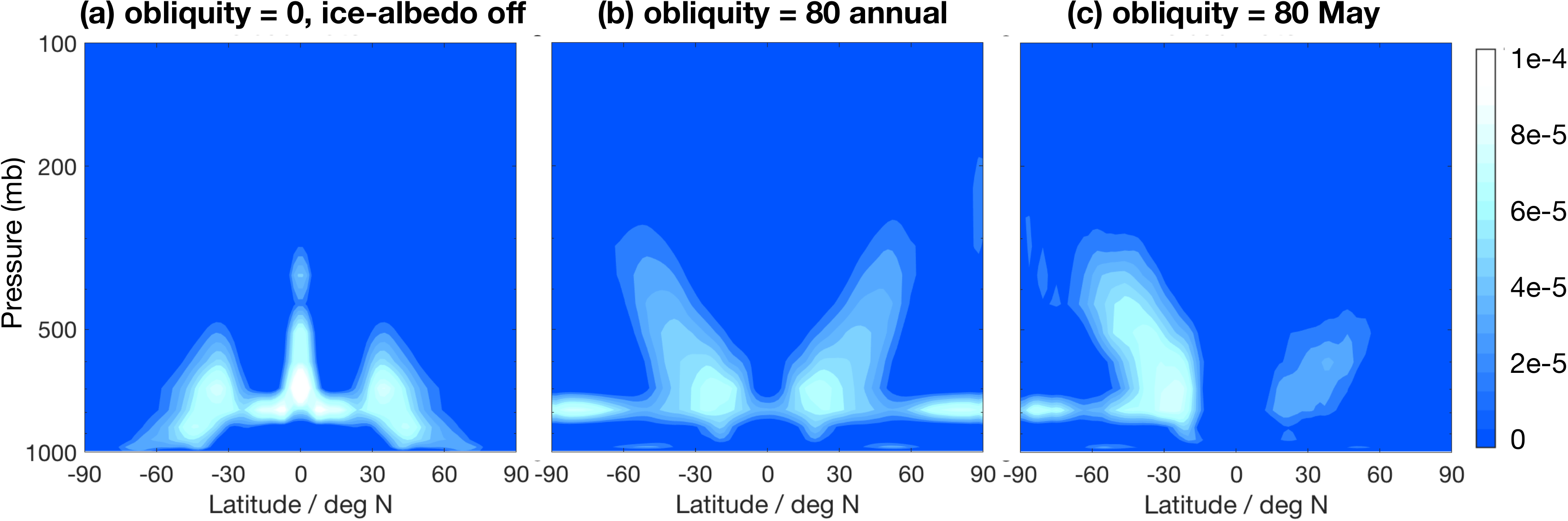}
 \caption{Climatological cloud water as a function of latitude and pressure. From left to right, shown are (a) annual mean of the low obliquity experiment with ice-albedo feedback turned off, (b) annual mean of the high obliquity experiment, and (c) May climatology of the high obliquity experiment.}
 \label{fig:cloud-water}
\end{figure*}

Cloud water (liquid+ice) distribution is shown in Fig.~\ref{fig:cloud-water}. For the low obliquity experiment, clouds are thickest at the equator year round (there is no seasonal cycle), giving rise to the high albedo at the equator (Fig.~\ref{fig:OLR-albedo}b,c). Clouds are collocated with the insolation maximum, effectively shielding the surface from solar radiation where it is strongest. Conversely, in the high obliquity experiment, annual-mean cloud thickness is greatest between 20-50N/S, which is offset from the insolation maximum at the high latitudes. As a result, the solar radiation reflection becomes less efficient. The global planetary albedo reaches its minimum during May and Nov in the high obliquity setup, and therefore we also show the May climatology of cloud water in Fig.~\ref{fig:cloud-water}(c). With the substellar point already moved to 54N, the SH is still warmer than the NH due to the heat stored in the ocean, leading to much higher cloud coverage (not shown) and cloud water concentration there. Although clouds do form, they are located mostly in the SH which receives little sunlight during that time of the year, meaning that the cloud reflection tends to be inefficient due to the seasonal lag between clouds and the sun under high obliquity.

While not directly related to the albedo, an interesting observation is that the cloud distribution tilts in opposite directions in the high and low obliquity experiments. We speculate this is because the isentropes tilt upward (downward) when approaching the poles under low (high) obliquity, and clouds form as they cross isentropes. Also, although the two poles are the hottest regions under high obliquity, only shallow clouds form there. This is possibly due to the lack of large scale upward motion there. As suggested by \citet{Faulk-Mitchell-Bordoni-2017:effects}, the solstice Hadley cell will be constrained in the tropics even with the highest temperature located at the poles.

The above diagnosis suggests that the lag between clouds and the sun may be the reason for the relatively warmer climate under high obliquity. Two factors are required for this mechanism to work: the cloud radiation effects and the seasonal variation. We examine the above hypothesis here. We first turn off the cloud radiation effects by setting the clouds' optical depth and single scattering albedo to zero. As shown in Fig.~\ref{fig:Ts-lat-feedbacks}(c), the 9K global-mean annual-mean surface temperature difference reduces to 1.6K, which may be due to either the different strength of water vapor greenhouse effect or the nonlinear relationship between the OLR and the surface temperature. The near vanishing of the temperature difference between high and low obliquity planets indicates that cloud radiation effects are necessary to create such a difference.

We then turn off the other factor, the seasonal variation, by applying the annual mean insolation at each latitude throughout the year. This helps us distinguish between having a low cloud coverage on annual average and having clouds offset from the peak of the solar radiation due to the seasonal variation.
Turning off the seasonal cycle, the global-annual mean surface temperature under high obliquity becomes even colder than that under low obliquity (Fig.~\ref{fig:Ts-lat-feedbacks}d), suggesting that the season-induced cloud-sun offset is the key mechanism.

\section{Conclusions}
\label{sec:conclusions}

We validated the relative warm climate on high obliquity planets at varying insolation, and then tried to understand the mechanisms through a series of feedback suppression experiments. The roles played by ice-albedo feedback and cloud radiation feedback and the seasonal variation were studied by sequentially turning them off in the model.

With gradually increased insolation (meaning we explore the cold branch), the high obliquity planets were shown to always be warmer than the low obliquity equivalents by at least 10 K. Under 1365 W/m$^2$ insolation (same as present-day Earth although without CO$_2$), ice reflection can explain a significant amount of the temperature contrast, particularly in the warm branch. However, there is still a 9K temperature contrast, when the ice-albedo feedback is switched off \citep[consistent with][ and our high insolation experiments]{Nowajewski-Rojas-Rojo-et-al-2018:atmospheric}, indicating that the relative warmness under high obliquity holds with and without ice. The cause of the temperature contrast was shown to be the inefficient cloud reflection under high obliquity. The ocean heat inertia creates a lag between the maximum surface temperature and the maximum solar radiation, causing the clouds to form more on the dark side, and reducing the cloud reflection even more.

We here focused on the obliquity effects, while the role of other parameters, e.g., atmospheric composition, rotation rate, mixed layer depth etc, are not explored. In particular, a surface heat inertia that is large enough to create lag between the sun and the cloud formation is crucial for our mechanism to work. This condition is likely to be satisfied if the surface is covered by liquid, in particular, water. However, the minimum of water required for this mechanism to function is still unclear, and therefore requires future study.


\acknowledgments
The author thanks Prof. Eli Tziperman and Prof. Ming Cai for insightful and helpful discussions. This work was supported by NASA Habitable Worlds program (grant FP062796-A) and NSF climate dynamics AGS-1622985. We would like to acknowledge high-performance computing support from Cheyenne provided by NCAR's Computational and Information Systems Laboratory, sponsored by the National Science Foundation. The relevant model outputs are archived in https://www.dropbox.com/sh/diu8y2r36nqsomc/AAC\_cIRPf14VdJs2spaO9MHwa?dl=0.

\end{document}